\documentclass{PoS}
\usepackage{epsfig}
\usepackage{amsmath}
\usepackage{amsfonts}
\usepackage{amssymb}
\newcommand{\beq}{\begin{equation}}
\newcommand{\eeq}{\end{equation}}
\newcommand{\bea}{\begin{eqnarray}}
\newcommand{\eea}{\end{eqnarray}}
\newcommand{\D}{\displaystyle}

\newcommand{\bx}{{\bf x}}
\newcommand{\tr}{{\rm tr}}

\newcommand{\re}{{\rm Re}}

\newcommand{\vev}[1]{\bigl\langle #1 \bigr\rangle}
\newcommand{\bpsi}{\bar{\psi}}

\title{Cluster expansions and chiral symmetry at large density in 2-color QCD }

\ShortTitle{Cluster expansions and chiral symmetry}

\author{\speaker{E. T. Tomboulis}
\\
        Dept. of Physics and Astronomy, University of California, Los Angeles\\
        Los Angeles, CA 90095, USA\\
        E-mail: \email{tomboulis@physics.ucla.edu}}

\abstract{$SU(N_c)$ lattice gauge theories with $N_f$ flavors of massless staggered fermions are considered 
at high quark chemical potential $\mu$ and any temperature $T$. In the strong coupling  regime (sufficiently small $\beta$) they have been shown to possess a chiral phase of intact global $U(N_f)\times U(N_f)$ symmetry. The proof is by cluster expansions which converge in the infinite volume limit.  Extension to weaker coupling does not appear feasible in the presence of complex fermion determinant. For theories with real determinant, however, such as 2-color QCD with fundamental fermions, or any $N_c$ with  even $N_f$ and adjoint fermions, 
such large $\mu$ cluster expansions can be used to show  chiral behavior of fermionic lattice observables at any gauge coupling. Unfortunately, this absence of color superfluidity/superconductivity at high $\mu$ appears to be a lattice artifact due to lattice saturation, a serious problem plaguing the standard finite density formalism on the lattice. Some possible ways of circumventing saturation are discussed.}

\FullConference{The 33rd International Symposium on Lattice Field Theory\\
		14 -18 July 2015\\
		Kobe International Conference Center, Kobe, Japan*}

\begin{document}

\section{Introduction}
A lot of effort has been devoted in recent years toward elucidating the expected rich structure of the QCD phase diagram. Still, away from a strip along the temperature axis at small density this phase diagram remains largely conjectural. This is due to our inability to perform simulations in Lattice Gauge Theory (LGT) due to the sign (complex fermion determinant) problem. With presently available simulation techniques we are basically restricted to $\mu/T \lesssim 1$. Even in cases with real determinant simulations at large $\mu$ with light fermions appear at least an order of magnitude more demanding than at zero density. It is this regime of high density and low temperature that is physically particularly interesting as it has been argued to engender, depending on the color and flavor content, a variety of color superconductivity/superfluidity  phases. 
In light of this state of affairs there have been many studies of finite density LGT 
at strong coupling which is amenable to a variety of techniques. 
Integrating out the gauge field in the strong coupling limit with staggered fermions results in a representation of the partition function in terms of monomers, dimers and baryon loops \cite{DMW}, or monomers, dimers and polymers 
\cite{KM}. The sign problem is partly evaded within this representation, thus allowing simulations 
\cite{DMW}-\cite{FdF}. Another approach is based on mean field investigations of effective actions obtained by retaining the leading terms in $1/d$ expansion in the spatial directions while leaving the timelike directions intact \cite{NFH}, \cite{KMOO}.  In all such investigations a transition to a chirally symmetric phase is found at some critical $\mu$. 
The existence of this phase for general $SU(N_c)$ at strong coupling was proven in \cite{T1}, \cite{T2} by means of a cluster expansion shown to converge for large $\mu$ in the infinite volume limit. Such large $\mu$ strong coupling cluster expansions are reviewed in section \ref{sc} below. We then  proceed to show how they can be used to extract information for 
all couplings in the case of LGT with real fermion determinant. We discuss the meaning of such lattice results in the last section.  

\section{Large $\mu$, strong coupling cluster expansion in $SU(N_c)$ LGT \label{sc}}
The lattice action is $S=S_g+ S_F$ where $S_g$ is the usual gauge field plaquette action and 
$S_F  =  \sum_{x,y} \bpsi(x){\bf M}_{x,y}(U)\psi(y) $ 
is the action for massless staggered fermions in the presence of quark chemical potential $\mu$.  
We take $N_f$ staggered fermions flavors (which corresponds to $4N_f$ continuum flavors).   
$S_F$ is then invariant under a $U(N_f)\times U(N_f)$ global chiral symmetry corresponding to independent rotations of fermions on even and odd sublattices. 

The basic idea behind the cluster expansion \cite{T1} is that 
the presence of a nonvanishing chemical potential in $S_F$ introduces an anisotropy between the spacelike and timelike directions. This can be exploited to set up a cluster expansion for large $\mu$. 
(There is an analogous anisotropy in the case of large $T$ that was used for the convergent expansion in \cite{TY} showing chiral symmetry restoration at high temperature.)
The expansion is generated by simply expanding the exponential of the space-like part of the action, $\exp \sum_{x,y} \bpsi(x){\bf M}_{x,y}^{(s)}(U)\psi(y)$ and carrying out the fermion integrations in the measure provided by the unexpanded  exponential of the timelike part of the action, $\exp \sum_{x,y} \bpsi(x){\bf M}_{x,y}^{(t)}(U)\psi(y)$, 
which depends on $\mu$. In other words, one performs a fermion spacelike hopping expansion with fermions connected in the time direction by propagators given by the $\mu$-dependent timelike part of the action. 
In the strong coupling limit, i.e., $\beta=0$, where the gauge action $S_g$ is absent, this constitutes the entire expansion \cite{T1}. It may be extended to finite strong coupling, i.e. small $\beta$, by combining this fermionic expansion with the usual strong coupling plaquette expansion \cite{T2}. The latter is obtained by 
expanding the exponential of the plaquette gauge field action in characters:
\beq
\prod_p \exp\{{\beta\over N_c }\re \,\tr U_p \}   =  a_0(\beta)^{|\Lambda|} \,\prod_p  \Big[ 1 + \sum_{j\not= 0} d_j c_j(\beta) \chi_j(U_p) \Big]  
    \equiv     a_0(\beta)^{|\Lambda}| \,\prod_p \Big[ 1 + f_p(U_p) \Big] \label{chexp2}
  \eeq
and expanding in powers of $f_p$'s. 
The diagrammatics of these expansions are explained in \cite{T1}, \cite{T2}. 
In the  Polyakov gauge where all bond variables $U_0(\tau, \bx)$ are chosen to be independent of $\tau$, i.e. 
\beq
 U_0(\tau, \bx) =  {\rm diag} (e^{i\theta_1(\bx)/L}, e^{i\theta_2(\bx)/L}, \cdots, e^{i\theta_{N_c}(\bx)/L} ) 
\equiv  \exp (i\Theta(\bx)/L)  \, , \label{polg}
\eeq 
explicit evaluation of the timelike fermion propagator for propagation from $\tau^\prime$ to $\tau$ gives \cite{T1}:
\bea
C(\tau - \tau^\prime, \Theta(\bx))_{ai,bj} &  = &   \delta_{ab}\delta_{ij} [1 - (-1)^{\D (\tau-\tau^\prime)}] 
\,  {  e^{\D -i\theta_a(\bx) (\tau-\tau^\prime)/L}  \over 
1 +e^{\D -i\theta_a(\bx) }  e^{ \D - \mu L}  } \, e^{\D - \mu (\tau - \tau^\prime)} \; ,  \nonumber \\
& & \qquad \qquad   \mbox{for} \quad (\tau-\tau^\prime) > 0, \quad \mu >0  \label{prop1a}
\eea
\bea 
 C(\tau - \tau^\prime, \Theta(\bx))_{ai,bj} 
& = & \! -\delta_{ab}\delta_{ij} [1 - (-1)^{\D |\tau-\tau^\prime|}] \,  
{e^{\D -i\theta_a(\bx)[1 -  |\tau-\tau^\prime|/L]} 
 \over  1 +e^{\D -i\theta_a(\bx) } e^{\D  - \mu L}  } \,
 e^{ \D - |\mu| [L -  |\tau - \tau^\prime|]}  \; , \nonumber \\
& & \qquad \qquad   \mbox{for} \quad (\tau-\tau^\prime) < 0, \quad \mu >0  \label{prop1b}
\eea
Note that, for $\tau^\prime > \tau$, propagating backward in time from $\tau^\prime$ to $\tau$ is equivalent to propagating forward from $\tau^\prime$ winding around the periodic time direction  to $\tau$.  
For $\mu < 0$, i.e., for nonvanishing antiquark chemical potential, a physically distinct situation, 
 replace $\theta_a(\bx)$ by $ -\theta_a(\bx)$,  and reverse the sign condition on $(\tau - \tau^\prime)$ in (\ref{prop1a}) and  (\ref{prop1b}). 

\begin{figure}
\begin{center}
\includegraphics[width=10cm]{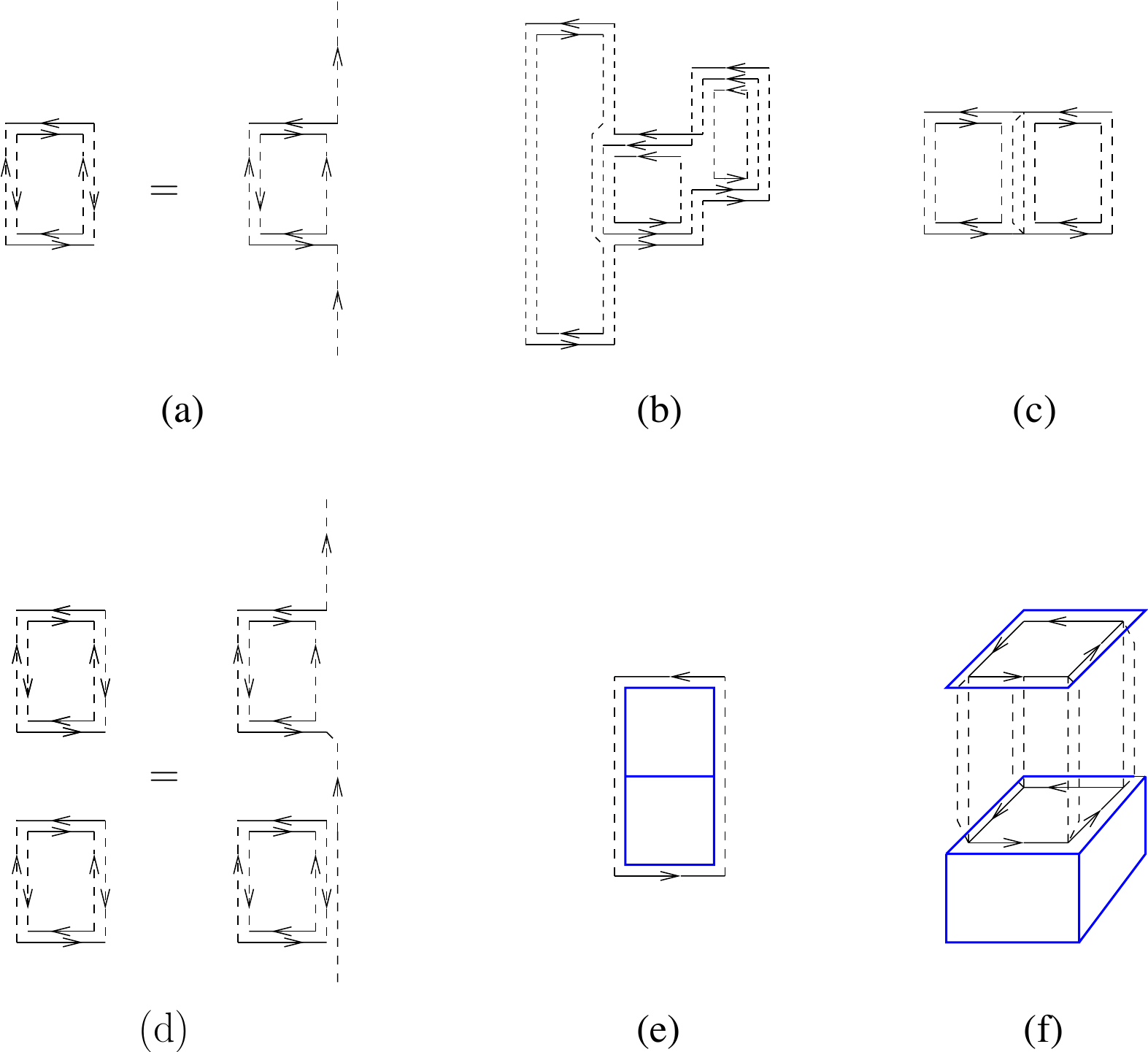} 
\end{center}
\caption{Some diagrams in the expansion: (a) - (d) diagrams involving only fermions (as in $\beta=0$); 
(e) - (f) diagrams including gauge field plaquettes. Directed lines represent fermion hopping spacelike links, broken lines represent timelike propagators (their direction shown for clarity only in (a) and (d)), gauge field plaquettes shown in solid blue lines.   \label{cpcr-hF1} }
\end{figure}

As seen from (\ref{prop1a}) - (\ref{prop1b}),  $C(\tau, \Theta(\bx))$  vanishes for even $\tau$. This is a consequence of the chiral invariance of the action. The other salient property of $C(\tau, \Theta(\bx))$ is its 
exponential decay for nonvanishing $\mu$. 
These are the crucial properties for the convergence of the expansion. Some typical diagrams are shown in Fig. \ref{cpcr-hF1}.  
The expansion can be shown to converge in the large volume limit for sufficiently large $\mu$ and small $\beta$ \cite{T1}, \cite{T2}. 

A consequence of such convergence is that the 
expectation of any local chirally non-invariant fermion operator ${\cal O}(x)$, e.g., $\bpsi(x)\psi(x)$ or diquark operators, vanishes identically  term by term in the expansion by the  invariance of the measure. 
Correlation functions $\vev{{\cal O}(x) {\cal O}(y)}$ then can receive 
non-vanishing contributions only from diagrams intersecting  both sites $x$ and $y$. 
A straightforward consequence of this fact is that: 
\[ 
\left|\vev{{\cal O}(x) {\cal O}(y)}\right|  < {\rm C_0} 
{\rm C}^{- |x-y|}  \, , \label{corrbound}
\]
where $C_0, C$ are space-dimension-dependent constants and $|x-y|$ is the minimum number of bonds connecting the two sites. In other words, there is {\it clustering} of 2-point (and all higher) correlations: 
the global $U(N_f)\times U(N_f)$ symmetry is intact for sufficiently large $\mu$, and small $\beta$.

\section{Theories with real fermion determinant} 
Can we extend this expansion setup to larger regimes of the gauge coupling? 
Write the partition function in the form of a gauge field integration over a fermion partition function $Z_F(U)$: 
\beq  Z=\int D U\,  e^{S_g(U)} Z_F(U)  \qquad \mbox{with} \qquad 
Z_F(U) = \int D\bpsi D\psi \,  e^{S_F(U)}= {\rm  Det}\,{\bf M}(U)  \label{PF1} \, . 
\eeq 
The expectation of a general fermionic operator $O[\bpsi, \psi]$ may then be expressed in the form: 
\beq
\vev{O}  
= \int d\nu(U) \, \vev{O}_F\!(U)  \; , \label{Fexp1}
\eeq
where 
\beq
\vev{O}_F\!(U) = {1\over Z_F(U)} 
\int D\bpsi D\psi \,  e^{S_F(U)} O[\bpsi, \psi]  \,     \label{Fexp2}
\eeq
is its expectation in the fermionic measure in the background of the gauge field and 
 \beq 
 d\nu(U) \equiv  {dU \over Z} e^{S_g(U)}\,{\rm  Det}{\bf M}(U) \label{PF2} 
 \eeq
 is the 
(normalized) full effective gauge field measure at coupling $\beta$.

Now, one may expand the fermion expectation $\vev{O}_F (U)$ given by (\ref{Fexp2})   
in the same type of expansion as in the previous section. 
This  expansion for 
$\vev{O}_F (U)$, in generic gauge field background $U$ and for operators $O$ of bounded support   
and spatial dimension $d \geq 1$, converges absolutely, and uniformly in the spatial lattice size, at any  temperature $T$ for  sufficiently large $\mu$.  The result holds for any choice of $N_c$, $N_f$. 
The proof, and associated estimates, proceed  as in the strong coupling case above except that it is actually {\it simpler} since no integration over the gauge field is involved engendering additional connectivity among diagrams.

Expansion of $\vev{O}_F (U)$ leads to expansion of 
\beq \vev{O}   
= \int d\nu(U) \, \vev{O}_F\!(U)  \; . \label{Fexp3}
\eeq
Does this expansion converge? 
Convergence of the $\vev{O}_F (U)$ expansion implies 
\beq
|\vev{O}_F (U)| <  C_{O}  \, ,  \label{upperb1}
\eeq
where $C_O$ is a constant, which is observable- and spacetime-dimension-dependent, but 
independent of the background $U$ and the spatial lattice volume. 
Absolute convergence of the expansion (\ref{Fexp3}) for $\vev{O}$ now follows from the absolute convergence of the expansion for  $\vev{O}_F\!(U)$ {\it provided} the measure $d\nu$ is real and positive:   
\beq
| \vev{O} |   
= \int d\nu(U) \, | \vev{O}_F\!(U) | <   C_{O} \; . \label{Fexp4}
\eeq
$d\nu(U)$ is real positive if the fermion determinant ${\rm  Det}{\bf M}(U)$ is real positive. This is the case for $N_c=2$ and fundamental rep. fermions; or any $N_c$, even $N_f$ and adjoint fermions.   
In the case of $N_c=2$ and fundamental fermions one has a pseudo-real representation 
with gauge field matrices satisfying  $\tau_2 U \tau_2 = U^*$. In the case of general $SU(N_c)$ and adjoint fermions one has a real representation
and the $U$'s represented by real orthogonal matrices.  In both cases then ${\rm Det} {\bf M}$ is real: in the two-color and fundamental fermion case ${\rm Det} {\bf M} ={\rm Det} \tau_2{\bf M}\tau_2= {\rm Det} {\bf M}^*$; in the general $N_c$ and adjoint fermion case ${\rm Det} {\bf M}$ is manifestly real. 
Furthermore, at $\mu=0$ the $U(N_f)\times U(N_f)$ symmetry of staggered fermions is enlarged to $U(2N_f)$, 
which has interesting consequences for spontaneous symmetry breaking, cf. \cite{H1}.  At low $T$, a  sequence of a chiral condensate phase, followed by a diquark condensate phase, followed by a chiral symmetry restored phase is  expected with increasing $\mu$, as found in mean field computations, cf. \cite{DMW}, \cite{NFH}. Extensive simulations of the 2-color theory at finite $\mu$ with Dirac fermions have been carried out in \cite{H2}. Unless close to the continuum limit, however, Dirac fermions do not possess any well-defined chiral properties and cannot be compared to staggered fermions in a meaningfully way. 

Operators $O$ of interest here would be the usual chiral condensate order parameter $\D  O_{\bar{q}q}= \bpsi(x)\psi(x)$, as well as the 
$N_c=2$ fundamental fermions diquark condensate: 
\beq  
O_{qq} = {1\over 2} \left[ \psi^T(x) \tau_2 \psi(x) + \bpsi(x) \tau_2 \bpsi^T(x) 
\right] \; . 
\eeq
For $SU(N_c)$ an adjoint fermions diquark condensate
\beq 
O_{qq}^{{\rm ad}\, ij\ldots} = {1\over 2} \epsilon^{ij\ldots kl}\left[ \psi^{k\,T}(x) \psi^l(x) + \bpsi^k(x) \bpsi^{l\,T}(x) 
\right]   \label{diqad}
\eeq
breaks $U(N_f)\times U(N_f) \longrightarrow SU(2)$ isospin; in particular, the condensate breaks 
$U_B(1)$. Such breaking of the global symmetries engenders superfluidity. 
An operator for color symmetry breaking condensate would be 
\beq 
O_{qq}^{a} =  {1\over 2} \left[ \psi^T(x) t^a \psi(x) + \bpsi(x) t^a \bpsi^T(x) 
\right] \, . \label{diqcolor}
\eeq
(\ref{diqcolor}) transforms in the adjoint representation of $SU(N_c)$, i.e. as a composite adjoint Higgs field, and its condensation would break the color symmetry, as well as the chiral symmetries, resulting into color superconductivity. Note that because it is an adjoint composite, any such phase would be separated from the unbroken confining phase by a true phase boundary.

 An immediate consequence of the convergence implied by (\ref{Fexp4}), however, is that, just as before (section \ref{sc}), within the convergence radius the $U(N_f)\times U(N_f)$ symmetry is preserved. 
Indeed, the expectation of any local fermion operator non-invariant under this symmetry, such as $\D  O_{\bar{q}q}$, $O_{qq}^{{\rm ad}\, ij\ldots}$ or   $O_{qq}^{a} $  above,         
vanishes identically term by term in our expansion by the invariance of the measure. Equivalently, all 2-point and higher correlation functions of such operators are seen to cluster exponentially, i.e., there is no spontaneous breaking of the global $U(N_f)\times U(N_f)$ symmetry.  The result holds for all gauge couplings $\beta$ and all temperatures $T$ at sufficiently large $\mu$. 
As a consequence no superfluidity and/or color superconductivity phase involving breaking of (any part of) these global symmetries occurs 
at high $\mu$. Rather what may be called a ``quarkyonic" phase \cite{McLP} with intact chiral symmetry obtains at low $T$.

\section{Lattice saturation - Discussion} 
We saw that at sufficiently large quark chemical potential and sufficiently large gauge coupling the $U(N_f) \times U(N_f)$ global symmetry of the $SU(N_c)$ LGT with $N_f$ flavors of massless staggered fermions is intact. This is an exact lattice result obtained by cluster expansions converging in the large volume limit. It accords with a large number of previous simulation and mean-field/analytical studies, mostly for $N_c=3,2$, \cite{DMW}-\cite{KMOO} which find a transition to such a phase. Furthermore, in the case of LGT with real fermion determinant we saw how information obtained from  these fermionic cluster expansions  can be used to extend this result to all couplings. The crucial question, of course, is what relation these lattice results bear to the continuum massless theory. 

Unfortunately, the immediate answer appears to be that they are of little direct relevance. This is because they seem to be largely determined by the onset of lattice saturation. Lattice saturation, i.e., every lattice site being occupied by the maximum number of fermions allowed by the Pauli principle, can be a real effect on a physical lattice, as observed in certain condensed matter systems;  but it is a regularization artifact in the LGT context.   
Once saturation sets in no condensates can form. Computation of the quark number density  within our expansion indeed shows that saturation is present; the density is at its maximum per site at $T=0$ deviating only by small exponential corrections at low $T$. The saturation effect at strong coupling  sets in immediately upon the transition to the chirally symmetric phase. An earlier discussion of this was given in \cite{ADiCGL}. 
This raises the question of whether the $T=0$ chiral transition at strong coupling seen in simulations, most recently in \cite{FdF}, reflects the eminent set-in of lattice saturation rather than the true location of the (expected) transition. To explore such questions would require having some control over the onset of saturation.

In the usual finite density lattice formalism, employed here and previous cited studies,  
a chemical potential $\mu$ is introduced on the timelike links uniformly throughout the lattice. 
This inexorably leads to saturation once $\mu$ becomes large enough. A possible way of avoiding this is to introduce spatially variable $\mu$, and, in particular, a ``thinned-out" distribution obtained by setting $\mu$ to a lower or negative value on a subset of the timelike bonds.
One may, e.g., partition the spatial lattice into cubes of some fixed size and introduce $\bar{\mu} < 0$ (antiquark chemical potential) on the bonds of the timelike fiber(s) extending from one (or more) site(s) in each cube, while having  quark chemical potential $\mu>0$ on the rest of the timelike links.  
One may thus achieve any mean density (net particle number per unit volume) by adjusting $\mu$ versus $\bar{\mu}$, while allowing local particle number fluctuations which thwart complete saturation.  
It is not hard to see that such schemes will generally upset the convergence of the expansions above, and 
may necessitate some series repackaging or resummation leading to a different physical 
picture.\footnote{The chiral symmetry transition in going from small $N_f/N_c$ to large $N_f/N_c$ at strong coupling may provide an analogous example of such resummations necessitated by a change in physical parameters and completely altering the physical outcome \cite{T3}.}  Such extensions are currently under investigation.

\end{document}